\documentclass[11pt]{article}

\usepackage{thmtools}
\usepackage{amsthm}
\usepackage{amsmath,xcolor,amssymb,enumitem}

\usepackage{thm-restate}
\usepackage{hyperref}
\hypersetup{colorlinks,linkcolor=blue,citecolor=blue}
\usepackage{cleveref}
\usepackage{fullpage,palatino,setspace}
\usepackage{graphicx} 
\usepackage{mathtools,dsfont}
\usepackage{mathrsfs}
\newtheorem{theorem}{Theorem}[section]
\newtheorem{fact}[theorem]{Fact}
\newtheorem{definition}[theorem]{Definition}
\newtheorem{lemma}[theorem]{Lemma}
\newtheorem{conj}{Conjecture}

\newcommand{\bool}{{\{0,1\}}}
\newcommand{\sbool}{\{-1,1\}}
\newcommand{\boolfn}[1]{\ensuremath {\sbool}^{#1} \rightarrow \sbool}
\newcommand{\cfn}[1]{\ensuremath {\sbool}^{#1} \rightarrow \complex}

\newcommand{\bfn}[1]{\ensuremath {\sbool}^{#1} \rightarrow \bool}

\DeclareMathOperator{\ent}{\ensuremath \mathbb{H}}
\DeclareMathOperator{\ente}{\ensuremath \mathcal{E}}
\DeclareMathOperator{\bor}{\ensuremath \mathsf{OR}}
\DeclareMathOperator{\band}{\ensuremath \mathsf{AND}}
\DeclareMathOperator{\maj}{\ensuremath \mathsf{Maj}}
\DeclareMathOperator{\E}{\ensuremath \mathbb{E}}

\newcommand{\fc}[2]{\ensuremath {\widehat{#1}(#2)}}
\newcommand{\fcsq}[2]{\ensuremath {{|\widehat{#1}}(#2)|^2}}
\newcommand{\reals}{\ensuremath {\mathbb{R}}}
\newcommand{\complex}{\ensuremath{\mathbb{C}}}
\newcommand{\codim}{\ensuremath {\mathsf{co\text{-}dim}}}

\newcommand{\s}{\ensuremath {\mathsf{s}}}
\newcommand{\as}{\ensuremath {\mathsf{as}}}
\renewcommand{\deg}{\ensuremath {\mathsf{deg}}}
\newcommand{\avgcj}{\ensuremath {\mathsf{avgdeg}_{\mathsf{cj}}}}
\newcommand{\certificate}{\ensuremath {\mathsf{C}}}
\newcommand{\ac}{\ensuremath {\mathsf{aC}}}

\newcommand{\rdt}{\ensuremath {\mathsf{R}}}

\newcommand{\aUC}{\ensuremath {\mathsf{aUC}}}
\newcommand{\UC}{\ensuremath {\mathsf{UC}}}
\newcommand{\dnf}{\ensuremath {\mathsf{DNF}}}
\newcommand{\Inf}{\ensuremath {\mathsf{Inf}}}

\newcommand{\Var}{\ensuremath {\mathsf{Var}}}
\newcommand{\pmset}[1]{\{-1,1\}^{#1}} 

\newcommand{\conjugate}[1]{\overline{#1}}

\newcommand{\Qcc}{\mathsf{Q}^{*,\mathrm{cc}}}
\newcommand{\norm}[1]{\left\lVert#1\right\rVert}
\newcommand{\Frob}{\mathrm F}
\newcommand{\tn}{\mathrm{tr}}

\title{Fourier Entropy Bounds via Approximations}
\author{Manaswi Paraashar\footnote{Indian Institute of Technology Hyderabad. Email: {\tt manaswi@cse.iith.ac.in}}
 \and
 Nitin Saurabh\footnote{Indian Institute of Technology Hyderabad. Email: {\tt nitin@cse.iith.ac.in}}
}
\date{}

\begin{document}

\maketitle

\begin{abstract}
We establish upper bounds on the Shannon entropy $\mathbb H(\widehat f^{\,2})$ of the Fourier distribution of Boolean functions. We prove that
\[
\mathbb H(\widehat f^{\,2})
=O\!\left(\log\bigl(1+\|\widehat f\|_{1,1/3}\bigr)\right),
\]
where \(\|\widehat f\|_{1,1/3}\) denotes the minimum spectral norm of a pointwise \(1/3\)-approximant to \(f\). This implies \(\mathbb H(\widehat f^{\,2})=O(\mathsf R^\oplus(f))\), where \(\mathsf R^\oplus(f)\) is bounded-error randomized parity-query complexity. We also bound Fourier entropy by the smaller of the two one-sided average unambiguous parity-certificate complexities. More generally, for any exact representation of indicator of $f$ as \(\sum_C w_C\mathbf1_C\), with nonnegative weights and affine subspaces \(C\), we prove
\[
\mathbb H(\widehat f^{\,2})
=O\!\left(\sum_C w_C\,2^{-k_C}k_C\right),
\qquad k_C=\operatorname{codim}(C).
\]
The quantity on the right is the average conical junta degree of the representation. These results strengthen earlier bounds of \cite{ACKSW21} by extending approximation-based entropy estimates to Shannon entropy and allowing one-sided and overlapping exact representations. 
The proofs use an entropy bound from a sequence of increasingly accurate approximants of $f$ with \emph{small} spectral norm.

Our techniques also find an application in quantum communication complexity, where it improves a lower bound for the bounded-error quantum communication complexity of XOR functions.

\end{abstract}

\newpage

\section{Introduction}
\label{sec:intro}
The Fourier Entropy-Influence Conjecture (FEIC) of Friedgut and
Kalai~\cite{FK96} proposes a fundamental relation between the Fourier
spectrum of a Boolean function and its sensitivity to changes in the
input. For $f\colon\boolfn{n}$, Parseval's identity gives
$\sum_{S\subseteq[n]}\fcsq{f}{S}=1$, so the squared Fourier coefficients
define a probability distribution on subsets of $[n]$. Two natural
quantities associated with this distribution are its Shannon entropy $(\ent(\widehat{f}^2))$ 
and the expected size of a set drawn from it $(\Inf(f))$:
\begin{align*}
  \ent(\widehat{f}^2)
  &=\sum_{S\subseteq[n]}\fcsq{f}{S}\log\frac{1}{\fcsq{f}{S}},
  &\Inf(f)&=\sum_{S\subseteq[n]}|S|\fcsq{f}{S}.
\end{align*}
The latter quantity is the total influence $\Inf(f)$ of $f$, or equivalently,
its average sensitivity. Throughout the paper, logarithms are to base
two unless otherwise indicated, and $0\log(1/0)=0$.

\begin{conj}[Fourier Entropy-Influence Conjecture~\cite{FK96}]
\label{conj:fei}
There is a universal constant $C>0$ such that every Boolean function
$f\colon\boolfn{n}$ satisfies
\[
  \ent(\widehat{f}^2)\leq C\cdot\Inf(f).
\]
\end{conj}

The conjecture asserts that a Boolean function with small total
influence cannot have its Fourier weight spread over too many
coefficients in an information-theoretic sense. Its original
motivation came from threshold phenomena for monotone graph
properties~\cite{FK96}. A positive resolution would strengthen
sharp-threshold estimates \cite{BK97,KKLMS19} and imply the Kahn--Kalai--Linial
theorem~\cite{KKL88}; see~\cite{OWZ11,KKLMS19,ACKSW21} for these implications. 
The conjecture also connects the analysis of Boolean functions with
learning theory through sparse approximation in the Fourier basis.

A central example of this connection is Mansour's
conjecture~\cite{Mansour95}. A $t$-term $\dnf$ is a disjunction of at
most $t$ conjunctions of literals. Although its Fourier expansion can
have exponentially many non-zero coefficients, the conjecture predicts
that a polynomial number of them captures almost all of its Fourier
weight.

\begin{conj}[Mansour's conjecture~\cite{Mansour95}]
\label{conj:mansour}
Let $f\colon\boolfn{n}$ be computed by a $t$-term $\dnf$, where $t\geq2$.
For every $0<\varepsilon<1$, there is a family
$\mathcal{S}\subseteq2^{[n]}$ such that
\[
  |\mathcal{S}|\leq t^{O(\log(1/\varepsilon))}
  \qquad\text{and}\qquad
  \sum_{S\in\mathcal{S}}\fcsq{f}{S}\geq1-\varepsilon.
\]
\end{conj}

By Parseval's identity, this is a statement about approximation by
sparse Fourier polynomials in squared $\ell_2$ distance. More generally,
an entropy bound $\ent(\widehat{f}^2)\leq H$ implies that
$2^{O(H/\varepsilon)}$ Fourier coefficients carry at least
$1-\varepsilon$ of the weight. Indeed, the coefficients whose squared
magnitude is below $2^{-H/\varepsilon}$ have total weight at most
$\varepsilon$. Consequently, FEIC would imply Mansour's conjecture
using the bound $\Inf(f)=O(\log t)$ for $t$-term
$\dnf$s~\cite{OWZ11,ACKSW21}. Combined with the query-learning
algorithm of Gopalan, Kalai, and Klivans~\cite{GKK08}, Mansour's
conjecture would give a polynomial-time algorithm for agnostically
learning polynomial-size $\dnf$s under the uniform distribution, with
query access and any fixed accuracy. This is a long-standing open question in computational learning theory \cite{gopalan:DNF}. 

One approach towards FEIC is to bound Fourier entropy by other
measures of Boolean function complexity. Chakraborty, Kulkarni, Lokam,
and Saurabh~\cite{CKLS16} obtained bounds in terms of average
decision-tree depth, including its parity-query analogue.
Arunachalam, Chakraborty, Kouck\'{y}, Saurabh, and de
Wolf~\cite{ACKSW21} strengthened these bounds to
$\ent(\widehat{f}^2)\leq2\aUC^{\oplus}(f)$, where
$\aUC^{\oplus}(f)$ is the average codimension of a monochromatic affine
subspace partition. They also bounded Fourier min-entropy by the
logarithm of approximate spectral norm and by randomized parity-query
complexity. Their corresponding bounds for R\'{e}nyi entropy of order
$1+\delta$ incur a factor that diverges as $\delta$ tends to zero, and
therefore do not give bounds on Shannon entropy. These results leave
two natural questions: can the approximation and randomized-query
bounds be extended to Shannon entropy, and can the partition bound
be expressed using only one output side or a nonnegative overlapping
representation?

\subsection{Our results}
\label{sec:our-results}

We answer both questions affirmatively. Our bounds extend the approximate spectral-norm and randomized-query results of \cite{ACKSW21} to Shannon entropy, and strengthen their partition bound to one-sided average unambiguous certificates and exact conical representations. 
Our proofs use two analytic ingredients: an entropy bound from a sequence of increasingly accurate approximants, and moment inequalities obtained by interpolation. 
The approximation method also gives a linear entropy lower bound for bounded-error, entanglement-assisted quantum communication. We state these results
below and then describe the general analytic techniques behind them.

\paragraph{Approximate spectral norm and randomized query complexity.}
Recall that $\|\widehat{f}\|_{1,\varepsilon}$ is the least spectral norm
of a polynomial $P$ satisfying $|f(x)-P(x)|\leq\varepsilon$ at every
input. The elementary inequality
$\ent(\widehat{f}^2)\leq2\log\|\widehat{f}\|_1$ bounds entropy by the
spectral norm of the exact function. We show that a uniform
approximation at any fixed error below one suffices. In particular,
Theorem~\ref{thm:approx-L1-norm} gives
\begin{equation}
\label{eq:intro-approx}
  \ent(\widehat{f}^2)
  =O\!\left(\log\bigl(1+\|\widehat{f}\|_{1,1/3}\bigr)\right).
\end{equation}
For any other fixed $0\leq\varepsilon<1$, the same statement holds
with a constant depending only on $\varepsilon$. Thus, the Fourier
distribution has \emph{small} Shannon entropy whenever $f$ admits an $\varepsilon$-approximating polynomial of \emph{small} spectral norm, 
even if the spectral norm of $f$ itself is much larger.

This yields an entropy bound in terms of bounded-error randomized
query complexity. Let $\rdt^{\oplus}(f)$ denote the minimum worst-case
number of parity queries needed to compute $f$, with error at most
$1/3$ on every input. Theorem~\ref{thm:rdt} states that
\begin{equation}
\label{eq:intro-rdt}
  \ent(\widehat{f}^2)=O\!\left(\rdt^{\oplus}(f)\right).
\end{equation}
In particular, $\ent(\widehat{f}^2)=O(\rdt(f))$ for ordinary randomized
query complexity. The expected output of a randomized parity-decision
tree is an approximating polynomial whose spectral norm is at most
$2^{\rdt^{\oplus}(f)}$; applying~\eqref{eq:intro-approx} at constant
error proves the bound. Equations~\eqref{eq:intro-approx}
and~\eqref{eq:intro-rdt} extend the corresponding results of \cite{ACKSW21}
from min-entropy and R\'{e}nyi orders greater than one to Shannon entropy.

\paragraph{One-sided average unambiguous certificates.}
For $b\in\bool$, consider a partition $\mathcal{U}_b$ of
$f^{-1}((-1)^b) \subseteq \sbool^n$ into affine subspaces. Its average cost is
\[
  \sum_{C\in\mathcal{U}_b}\codim(C)2^{-\codim(C)},
\]
and minimizing this quantity gives $\aUC_b^{\oplus}(f)$. The average
is taken over the entire uniform input space, with zero cost outside
the specified output side. Theorem~\ref{thm:ent-auc} gives
\begin{equation}
\label{eq:intro-auc}
  \ent(\widehat{f}^2)
  =O\!\left(\min\{\aUC_0^{\oplus}(f),\aUC_1^{\oplus}(f)\}\right).
\end{equation}
Restricting to subcube partitions gives the same conclusion for
ordinary one-sided unambiguous certificates. In particular, an
unambiguous $\dnf$ representation suffices to control the entropy by
the average width of its satisfied term, without constructing a
partition of the other output side. This improves the dependence on
the complexity measure in~\cite{ACKSW21}: their two-sided bound uses
$\aUC^{\oplus}(f)=\aUC_0^{\oplus}(f)+\aUC_1^{\oplus}(f)$, and their
one-sided extension includes an additional maximum-codimension term. 

\paragraph{Average conical junta degree.}
We extend the unambiguous certificate bound to exact non-negative representations
whose terms may overlap. In the usual definition, a conical junta is
a non-negative linear combination of conjunctions~\cite{GJ16}. We
also allow indicators of affine subspaces, and consider
representations
\[
  \frac{1-f}{2}=h=\sum_C w_C\mathbf{1}_C,
  \qquad w_C\geq0.
\]
The average conical junta degree of this representation is
\[
  \avgcj^{\oplus}(h)
  =\sum_Cw_C\codim(C)2^{-\codim(C)}.
\]
This is the uniform-input average of the local weighted degree.
For conjunctions, G\"o\"os and Jayram~\cite{GJ16} study the maximum
over inputs of that local quantity; our measure instead averages
over inputs. Minimizing over exact representations defines
$\avgcj^{\oplus}(f)$. Theorem~\ref{thm:ent-conical-junta} states that
\begin{equation}
\label{eq:intro-conical}
  \ent(\widehat{f}^2)
  =O\!\left(\min\{\avgcj^{\oplus}(f),\avgcj^{\oplus}(-f)\}\right).
\end{equation}
An unambiguous partition is a representation by disjoint indicators
with unit weights, so~\eqref{eq:intro-conical} generalizes
\eqref{eq:intro-auc}. The exact identity is essential: on the
represented output side the weights of the containing subspaces must
sum to one. An arbitrary $\dnf$ need not satisfy this condition, so
the result does not by itself give the entropy bound required for
Mansour's conjecture.

\paragraph{Quantum communication complexity.}
Our approximation method also bounds the entropy of squared singular
values. Let $M\in\{-1,1\}^{X\times Y}$ be a total sign matrix and let
$u,v$ be probability vectors on $X,Y$. The matrix
$A_{u,v}=D_{\sqrt u}MD_{\sqrt v}$ has squared Frobenius norm one.
Write $H_{u,v}(M)$ for the Shannon entropy of its squared singular
values. Theorem~\ref{thm:ent-qcc} shows that, for every fixed
$0\leq\varepsilon<1/2$,
\begin{equation}
\label{eq:intro-qcc}
  \Qcc_\varepsilon(M)\geq c_\varepsilon H_{u,v}(M),
\end{equation}
where arbitrary prior entanglement is allowed and $c_\varepsilon>0$
is independent of the matrix dimensions and the product weights.
For the XOR matrix $M_f(x,y)=f(x\oplus y)$, where $x\oplus y$ denotes
coordinatewise multiplication on $\pmset{n}$, uniform weights give
the Fourier distribution of $f$ as the squared singular-value
distribution. Consequently,
\[
  \Qcc_\varepsilon(M_f)
  =\Omega_\varepsilon\!\left(\ent(\widehat{f}^2)\right).
\]
Thus Fourier entropy provides a lower bound on bounded-error quantum
communication for XOR functions, even with shared entanglement.

\paragraph{Analytic ingredients.}
The approximation theorem, Theorem~\ref{thm:ent-approximants}, applies
to any $f\colon\cfn{n}$ with $\|f\|_2^2\leq1$. Suppose that for
every integer $j\geq1$ there is an approximant $P_j$ such that
\[
  \|f-P_j\|_2^2\leq2^{-j},
  \qquad
  \log\max\{1,\|\widehat{P_j}\|_1\}\leq CL(j+1)^a,
\]
where $a\geq0$, $C>0$, and $L\geq1$. Then
$\ente(\widehat{f}^2)=O_{a,C}(L)$, where $\ente$ denotes the entropy
functional without requiring the squared coefficients to sum to
one. The proof groups coefficients according to the first accuracy
at which they are well approximated. The approximation error bounds
the Fourier weight of the remaining groups, while the spectral norm
controls their entropy. Error reduction by majority supplies the required
sequence of approximants from a constant-error approximating polynomial to give \eqref{eq:intro-approx}. For
communication protocols, error reduction and the factorization-norm
estimate of Linial and Shraibman~\cite{LS07} supply analogous
approximants to the singular-value vector to give \eqref{eq:intro-qcc}. 

Our second ingredient is an interpolation inequality for sums. If
$f=\sum_i g_i$ with the nonzero $g_i$ pairwise orthogonal, set
$p_i=\|g_i\|_2^2$ and $L_i=\|\widehat{g_i}\|_1$.
Theorems~\ref{thm:moment-ineq-orthogonal-sums}
and~\ref{thm:ent-ineq-ortho} give
\begin{align*}
  \sum_S|\fc{f}{S}|^r
  &\leq\sum_iL_i^{2-r}p_i^{r-1}
  \qquad(1\leq r\leq2),\\
  \ente(\widehat{f}^2)
  &\leq2\sum_i p_i\log\frac{L_i}{p_i}.
\end{align*}
The first inequality follows from the Riesz--Thorin interpolation
theorem; differentiating at $r=2$, where equality holds, yields the
second. For the one-sided unambiguous representations, $g_i$'s are orthogonal, and hence Theorem~\ref{thm:ent-ineq-ortho} gives \eqref{eq:intro-auc}. Theorem~\ref{thm:moment-ineq-conical-junta} replaces
orthogonality by the condition that the Gram matrix
$\Gamma=(\langle g_i,g_j\rangle)_{i,j}$ has a diagonal majorant
$\Gamma\preceq\mathsf{diag}(a_i)$, where $a_i>0$. It gives the same moment bound
with $a_i$ in place of $p_i$, and the entropy bound when
$\sum_i a_i=\|f\|_2^2$. For nonnegative summands in an exact conical
representation, the row sums of $\Gamma$ provide such a majorant
with the required equality. This permits overlapping summands and
gives~\eqref{eq:intro-conical}.

\subsection{Prior work}
\label{sec:prior-work}
Several works have established FEIC for structured classes of Boolean
functions. O'Donnell, Wright, and Zhou~\cite{OWZ11} proved it for symmetric
functions and read-once decision trees. Chakraborty, Kulkarni,
Lokam, and Saurabh~\cite{CKLS16} and O'Donnell and
Tan~\cite{OT13} independently established it for read-once AND/OR formulas.
The latter work proves a composition theorem that also handles
read-once formulas over bounded-arity gates. Wan, Wright, and
Wu~\cite{www14} used a source-coding interpretation of FEIC to
reprove the composition theorem and to show that functions computed
by read-$k$ decision trees satisfy
$\ent(\widehat{f}^2)=O(k\Inf(f))$. Shalev~\cite{shalev2018:minentropy}
improved the dependence on $k$ to $O(\sqrt{k})$. In particular, these
results verify FEIC for every fixed read bound. In a very recent preprint, Shmalo~\cite{Shmalo26} establishes FEIC for every linear threshold function.  

There are also results for random functions. Klivans, Lee, and
Wan~\cite{KLW10} proved the FEI conjecture for random $\dnf$ formulas. 
Das, Pal, and Visavaliya~\cite{das:randomfei}
verified FEIC with high probability for a uniformly random Boolean
function. Chakraborty, Karmalkar, Kundu, Lokam, and
Saurabh~\cite{chakrabortyetal:feirandomltf} proved analogous results
for random linear threshold functions. Shalev~\cite{shalev2018:minentropy}
established FEIC when either the influence is \emph{too low} or the entropy is \emph{too high}. In the other direction, constructions of
Hod~\cite{hod:constantfei} show that any universal constant in FEIC
must exceed $6.45$, even for monotone functions. 

Kelman, Kindler, Lifshitz, Minzer, and Safra~\cite{KKLMS20} obtained
general bounds that bring Fourier concentration closer to the
prediction of FEIC. Among other results, they show 
\begin{align}
\label{eq:prior-kelman}
  \ent(\widehat{f}^2)
  =O\!\left(\sum_{S\subseteq[n]}
      |S|\log(|S|+1)\fcsq{f}{S}\right).
\end{align}
They also obtain bounds on large Fourier coefficients, with applications to learning and sharp thresholds. Han~\cite{Han25} proves the bound, 
\begin{align}
\label{eq:prior-han}
  \ent(\widehat{f}^2)
  =O\!\left(\Inf(f)+\sum_{i=1}^n
       \Inf_i(f)\log\frac{1}{\Inf_i(f)}\right).
\end{align}
The proof estimates Fourier moments iteratively and then obtains
entropy by differentiation. 
Our interpolation inequalities also pass from
moments to entropy via differentiation, but use orthogonal or non-negative decompositions in place of the individual influences. 

A different line of work bounds entropy by the complexity of a
description of the function. Chakraborty et al.~\cite{CKLS16}
bounded Fourier entropy in terms of average decision-tree depth,
including its parity-query analogue. Arunachalam, et al.~\cite{ACKSW21} strengthened these estimates to
\[
  \ent(\widehat{f}^2)\leq2\aUC^{\oplus}(f),
\]
where $\aUC^{\oplus}(f)$ is the minimum average codimension of a
monochromatic affine subspace partition. 
For ordinary subcubes, this gives a bound by average unambiguous certificate
complexity. 

\subsection{Comparison with previous work}
\label{sec:comparison}

The distinction between average cost and maximum cost is central to
the unambiguous certificate bounds. For a
partition $\mathcal{U}_b$, write $A_b$ for its average cost and
$d_b=\max_{C\in\mathcal{U}_b}\codim(C)$. The one-sided bound
in~\cite[Theorem~1.7]{ACKSW21} is
\[
  \ent(\widehat{f}^2)
  \leq2\bigl(A_b+\Pr[f\neq(-1)^b]d_b\bigr).
\]
Our bound is $O(A_b)$, with no dependence on $d_b$. The randomized
query bound addresses a different extension, allowing bounded error;
it should be distinguished from the earlier bounds using the average depth of an exact tree.

The approximation theorem controls the contribution of small Fourier
coefficients by combining approximants at all accuracies. It
therefore bounds the full Shannon entropy.  
In contrast to the influence-based estimates~\eqref{eq:prior-kelman}
and~\eqref{eq:prior-han}, our hypotheses concern the spectral norms
of approximants or the average cost of an exact decomposition.

For communication complexity, Montanaro and Winter~\cite{MW07}
proved an exact-computation lower bound of $H_{u,v}(M)/2$ and a
bounded-error bound of $\Omega(H_{u,v}(M)/\log n)$ for $n$-bit inputs,
allowing product weights and prior entanglement. For uniform
$N\times N$ sign matrices, Linial and
Shraibman~\cite[Theorem~30]{LS07} obtained
\[
  \Qcc_{1/3}(M)
  \geq\Omega\!\left(\frac{E}{1+\log((\log N)/E)}\right),
  \qquad E=H_{u,v}(M)>0.
\]
Our bound~\eqref{eq:intro-qcc} removes this denominator. The
improvement is relevant when $E$ is small relative to $\log N$. 
We use the same protocol factorization estimate underlying the
Linial--Shraibman method. The improvement is in deriving a
dimension-independent entropy consequence. Our proof combines
approximations at all accuracies and avoids a dimension-dependent
loss. 

\paragraph{Organization.}
Preliminaries are in Section~\ref{sec:prelims}. Section~\ref{sec:approximants} proves the
approximation theorem (Theorem~\ref{thm:ent-approximants}) and its consequences for approximate spectral norm (Theorem~\ref{thm:approx-L1-norm}) and randomized queries (Theorem~\ref{thm:rdt}). Section~\ref{sec:moment-ineq} develops
the moment and entropy inequalities (Theorems \ref{thm:moment-ineq-orthogonal-sums}, \ref{thm:ent-ineq-ortho}, and \ref{thm:moment-ineq-conical-junta}) and applies them to one-sided
certificates (Theorem~\ref{thm:ent-auc}) and conical juntas (Theorem~\ref{thm:ent-conical-junta}). Section~\ref{sec:application-qcc}
proves the quantum communication lower bound (Theorem~\ref{thm:ent-qcc}).

\section{Preliminaries}
\label{sec:prelims}

We recall some definitions and basic facts from analysis of Boolean functions,
referring to~\cite{ODonnell-book, wolf:fouriersurvey} for more.
Consider the space of all functions from $\pmset{n}$ to $\complex$ equipped
with the inner product defined as
\begin{align*}
  \langle f,g\rangle \coloneqq \E_x[\conjugate{f(x)}g(x)] =
  \frac{1}{2^n}\sum_{x\in \pmset{n}} \conjugate{f(x)}g(x).
\end{align*}
For $S\subseteq [n]$, the character function
\(\chi_S\colon\pmset{n}\rightarrow \pmset{}\) is defined as
\(\chi_S(x) \coloneqq \prod_{i\in S}x_i\).
Then the set of character functions $\left\{\chi_S\right\}_{S\subseteq [n]}$
forms an orthonormal basis for the space of
all complex-valued functions on $\pmset{n}$.
Hence, every complex-valued function
$f \colon\pmset{n}\rightarrow \complex$ has a unique Fourier expansion
\begin{align*}
  f(x) = \sum_{S \subseteq [n]} \fc{f}{S} \chi_S(x).
\end{align*} 
The \emph{degree} of $f$, denoted $\deg(f)$, is defined as
\(\max\left\{|S|:\fc{f}{S}\neq 0\right\}\).
The \emph{spectral norm} $\|\widehat{f}\|_1$ of $f$ is defined to be
$\sum_S|\fc{f}{S}|$. The \emph{Fourier weight} of a function $f$
on a set $\mathcal{S}$ of coefficients is defined as
$\sum_{S\in \mathcal{S}} |\widehat{f}(S)|^2$.
For $0 \leq \varepsilon < 1$, the $\varepsilon$-\emph{approximate spectral norm} $\|\widehat{f}\|_{1,\varepsilon}$ of a Boolean function $f$ is defined as
\begin{align*}
  \|\widehat{f}\|_{1,\varepsilon}=\min\Big\{\sum_S|\widehat{p}(S)| : |p(x)-f(x)|\leq \varepsilon \text{ for every } x\in \pmset{n}\Big\}.
\end{align*}
We will use $\widehat{f}\colon\cfn{n}$ to denote the function that maps $S$ to its Fourier coefficient $\fc{f}{S}$. 
We note a well-known fact that follows from
the orthonormality of the character functions.
\begin{fact}[Plancherel's Theorem]
  \label{fact:plancherel}
  For any $f,g \colon \pmset{n} \to \complex$,  
  \begin{align*}\langle f,g\rangle = \E_x[\conjugate{f(x)}g(x)] = \sum_{S\subseteq [n]}\conjugate{\fc{f}{S}}\fc{g}{S}=\langle \widehat{f},\widehat{g}\rangle.\end{align*}
\end{fact}
In particular, if $f\colon\pmset{n}\rightarrow \pmset{}$ is
Boolean-valued and $g = f$,
we have Parseval's Identity \(\sum_{S}\fcsq{f}{S} = \E\left[|f(x)|^2\right]\),
which in turn equals $1$.
Hence $\sum_{S}\fcsq{f}{S}=1$ and we can view $\left\{\fcsq{f}{S}\right\}_S$
as a probability distribution, which allows us to discuss
the Fourier entropy and min-entropy of the distribution
$\left\{\fcsq{f}{S}\right\}_S$, defined as 
\begin{definition}
  \label{def:shannon-entropy}
  For a Boolean function $f\colon\pmset{n}\rightarrow \pmset{}$,
  its Fourier entropy (denoted $\ent({\widehat{f}}^2)$) and
  min-entropy (denoted $\ent_\infty(\widehat{f}^2)$) are 
  \begin{align*}
    \ent(\widehat{f}^2) \coloneqq \sum_{S \subseteq [n]}\fcsq{f}{S} \log \frac{1}{\fcsq{f}{S}}, \quad \mbox{and} \quad \ent_\infty(\widehat{f}^2) \coloneqq \min_{\substack{S \subseteq [n]:\\\widehat{f}(S)\neq 0}} \Big\{\log \frac{1}{\fcsq{f}{S}}\Big\}.
  \end{align*}
\end{definition}
Similarly, we can also define the R\'{e}nyi Fourier entropy.
\begin{definition}[R\'{e}nyi Fourier entropy]
  \label{def:renyi-entropy}
  For $f\colon\pmset{n}\rightarrow \pmset{}$, $\alpha \geq 0$
  and $\alpha \neq 1$, the R\'{e}nyi Fourier entropy of $f$
  of order $\alpha$ is defined as
  \begin{align*}
    \ent_{\alpha}(\widehat{f}^2) \coloneqq \frac{1}{1-\alpha}\log\left(\sum_{S\subseteq [n]}|\fc{f}{S}|^{2\alpha}\right).
  \end{align*}
\end{definition}
It is known that in the limit as $\alpha \to 1$, 
$\ent_{\alpha}(\widehat{f}^2)$ is the (Shannon) Fourier entropy
$\ent(\widehat{f}^2)$ (see~\cite[Chapter~17, Section~8]{CT91}) and
when $\alpha \to \infty$,
observe that $\ent_{\alpha}(\widehat{f}^2)$ converges to $\ent_{\infty}(\widehat{f}^2)$.
It is easily seen  that
\begin{align*}
  \ent_{\infty}(\widehat{f}^2) \leq \ent_1(\widehat{f}^2) \leq
  \ent_{\frac{1}{2}}(\widehat{f}^2) \leq \ent_0(\widehat{f}^2).
\end{align*}
For an arbitrary function $g\colon\cfn{n}$, we will define 
\begin{align}
    \ente(\widehat{g}^2) := \sum_{S\subseteq [n]} \fcsq{g}{S}\log\frac{1}{\fcsq{g}{S}}. 
\end{align}
Note that $\sum_S\fcsq{g}{S}$ need not equal $1$. 
In the special case of Boolean functions $f\colon\boolfn{n}$ we have the following relationship.
\begin{lemma}
\label{lem:ente-to-ent}
    Let $f\colon\boolfn{n}$ be any Boolean function and $g := (1+f)/2$ (or, $(1-f)/2$). Then, 
    \begin{align}
        \ent(\widehat{f}^2) = 4 \ente(\widehat{g}^2) - \varphi(p) \leq 4 \ente(\widehat{g}^2), 
    \end{align}
    where $p=\E[g]=\Pr[g=1]$ and \(\varphi(p) = 4 p^2\log\frac{1}{p^2} + 8p(1-p) - (1-2p)^2\log\frac{1}{(1-2p)^2} \). Moreover, $\varphi(p) \geq 0$ for all $0\leq p \leq 1$. 
\end{lemma}

For $f\colon\pmset{n}\rightarrow \pmset{}$,
the \emph{influence} of a coordinate $i\in [n]$, denoted $\Inf_i(f)$,
is defined as
\begin{align*}
  \Inf_i(f)=\Pr_{x \in \sbool^n} \left[ f(x) \neq f(x^{(i)})\right]
  = \E_x\Big[\Big(\frac{f(x)-f(x^{(i)})}{2}\Big)^2\Big],
\end{align*}
where the probability and expectation is taken according to
the uniform distribution on $\pmset{n}$ and
$x^{(i)}$ is $x$ with the $i$-th bit flipped.
The \emph{total influence} of $f$, denoted $\Inf(f)$, is 
\begin{align*}
  \Inf(f)=\sum_{i\in [n]} \Inf_i(f).
\end{align*}
In terms of the Fourier coefficients of $f$,
it can be shown, e.g.,~\cite{KKL88}, that
\(\Inf_i(f)=\sum_{S\ni i} \widehat{f}(S)^2\), and therefore 
\begin{align*}
  \Inf(f)=\sum_{S\subseteq [n]} |S|\widehat{f}(S)^2.
\end{align*}
The \emph{variance} of a real-valued function $f$ is given by
\(\Var(f):=\sum_{S\neq \emptyset} \widehat{f}(S)^2 = 1 -\fcsq{f}{\emptyset}\).
It easily follows that $\Var(f)\leq \Inf(f)$. 
We now introduce some basic complexity measures of Boolean functions
which we use often, referring to~\cite{buhrman&wolf:dectreesurvey} for more. 
\paragraph{Sensitivity}
For $x \in \pmset{n}$, the \emph{sensitivity} of $f$ at $x$,
denoted $\s_f(x)$, is defined to be the number of neighbors
$y$ of $x$ in the Boolean hypercube
(i.e., $y$ is obtained by flipping \emph{exactly} one bit of $x$)
such that $f(y)\neq f(x)$. The sensitivity $\s(f)$ of $f$ is
$\max_{x}\{\s_f(x)\}$. The \emph{average sensitivity} $\as(f)$ of $f$
is defined to be $\E_x[\s_f(x)]$. By the linearity of expectation observe that 
\begin{align*}
  \E_{x}\left[\s_f(x)\right]=\sum_{i=1}^n \Pr_x\left[f(x)\neq f(x^{(i)})\right]
  =\sum_{i=1}^n\Inf_i(f)=\Inf(f),
\end{align*}
so the average sensitivity of $f$ equals the total influence of $f$.
As a result, the FEI conjecture asks if
\(\ent(\hat{f}^2)\leq C\cdot  \as(f)\) for every Boolean function $f$.
\paragraph{Certificate complexity}
For $x \in \pmset{n}$,
the \emph{certificate complexity} of $f$ at $x$,
denoted $\certificate(f,x)$,
is the minimum number of bits in $x$ that needs to be fixed
to ensure that the value of $f$ is constant.
The certificate complexity $\certificate(f)$ of $f$ is
$\max_x\{\certificate(f,x)\}$.
The minimum certificate complexity of $f$ is
\(\certificate_{\min}(f)=\min_x\{\certificate(f,x)\}\).
The $0$-certificate complexity $\certificate^{0}(f)$ of $f$ is
$\max_{x : f(x) = 1}\{\certificate(f,x)\}$.
Similarly, the $1$-certificate complexity
$\certificate^{1}(f)$ of $f$ is
$\max_{x : f(x) = -1}\{\certificate^{1}(f,x)\}$.
Observe that for every $x\in \pmset{n}$,
$\s(f,x)\leq \certificate(f,x)$.
This gives $\s(f)\leq \certificate(f)$ and
$\as(f)\leq \ac(f)$ where $\ac(f)$ denotes
the \emph{average} certificate complexity of $f$.
As before, the average is taken with respect to
the uniform distribution on $\pmset{n}$. 
\paragraph{Parity-certificate complexity}
Analogously, we  define the \emph{parity-certificate complexity}
$\certificate^{\oplus}(f,x)$ of $f$ at $x$ as the minimum number of
parities on the input variables one has to fix in order to fix 
the value of $f$ at $x$, i.e., 
\begin{align*}
  \certificate^{\oplus}(f,x) \coloneqq \min\{\mathsf{co\text{-}dim}(H) \mid H \text{ is an affine subspace on which }f\text{ is constant and }x \in H \},
\end{align*} 
where $\mathsf{co\text{-}dim}(H)$ is
the \emph{co-dimension} of the affine subspace $H$.
It is easily seen that \(\certificate^{\oplus}(f,x) \leq \certificate(f,x)\).
We also define
\(\certificate^{\oplus}(f) \coloneqq \max_x\{\certificate^{\oplus}(f,x)\}\),
and
\(\certificate_{\min}^{\oplus}(f) \coloneqq \min_x\{\certificate^{\oplus}(f,x)\}\).
\paragraph{Unambiguous certificate complexity}
We now define the \emph{unambiguous certificate complexity} of $f$.
Let $\tau\colon [n]\rightarrow\{-1,1,\ast \}$ be a partial assignment.
We refer to
\begin{align*}
  S_\tau=\left\{x\in \pmset{n}: x_i=\tau(i) \text{ for every } i\in [n]\setminus \tau^{-1}(\ast)\right\}
\end{align*}
as the subcube generated by $\tau$.
We call $C \subseteq \pmset{n}$ a \emph{subcube} of $\pmset{n}$
if there exists a partial assignment $\tau$ such that
$C=S_\tau$ and the {co-dimension} of $C$ is
the number of bits fixed by $\tau$, i.e.,
\(\mathsf{co\text{-}dim}(C)=|\{i\in [n]:\tau(i)\neq \ast\}|\). 
A set of subcubes $\mathscr{C}= \{C_1,\ldots ,C_m\}$ \emph{partitions}
$\pmset{n}$ if the subcubes are disjoint and they cover $\pmset{n}$,
i.e., $C_i\cap C_j=\emptyset$ for $i\neq j$ and $\cup_{i} C_i=\pmset{n}$.  

An \emph{unambiguous certificate} $\mathcal{U}=\{C_1,\ldots,C_m\}$,
also referred to as a \emph{subcube partition},
is a set of subcubes partitioning $\pmset{n}$.
We say $\mathcal{U}$ computes a Boolean function
$f\colon\pmset{n}\rightarrow \pmset{}$ if $f$ is constant on each $C_i$
(i.e., $f(x)$ is the same for all $x\in C_i$).
For an unambiguous certificate $\mathcal{U}$,
the \emph{unambiguous certificate complexity} $\UC(\mathcal{U},x)$
on input $x$ equals $\mathsf{co\text{-}dim}(C_i)$
for the $C_i$ satisfying $x\in C_i$.
Define the \emph{average unambiguous certificate complexity} of $f$
with respect to $\mathcal{U}$ as
\begin{align*}
  \aUC(f,\mathcal{U}) \coloneqq \E_x[\UC(\mathcal{U},x)].
\end{align*}
Then, the \emph{average unambiguous certificate complexity} of $f$
is defined as  
\begin{align*}
  \aUC(f)\coloneqq \min_{\mathcal{U}} \aUC(f,\mathcal{U}),
\end{align*}
where the minimization is over all unambiguous certificates for $f$. 
We also define the one-sided version of these measures. Let $\mathcal{U}=\mathcal{U}_0 \sqcup \mathcal{U}_1$ be such that, for $b\in\bool$, $\mathcal{U}_b$ is an unambiguous $b$-certificate, i.e., $\mathcal{U}_b$ is a subcube partition of $f^{-1}((-1)^b)$. Then, define $\aUC(f,\mathcal{U}_b)=(1/2^n)\sum_{x\in \mathcal{U}_b}\UC(\mathcal{U}_b,x)$ and note that $\aUC(f,\mathcal{U})= \aUC(f,\mathcal{U}_0)+ \aUC(f,\mathcal{U}_1)$. We therefore define the \emph{average unambiguous b-certificate complexity} of $f$ as 
\begin{align*}
  \aUC_b(f)\coloneqq \min_{\mathcal{U}_b} \aUC(f,\mathcal{U}_b),
\end{align*}
where the minimization is over all unambiguous $b$-certificates $\mathcal{U}_b$ for $f$. Observe that $\aUC(f) = \aUC_0(f)+\aUC_1(f)$. 
Finally, the \emph{unambiguous certificate complexity of $f$} is
\begin{align*}
  \UC(f) \coloneqq \min_{\mathcal{U}} \max_x \UC(\mathcal{U},x).
\end{align*}
Note that since unambiguous certificates are more restricted
than general certificates, we have $\certificate(f)\leq \UC(f)$.

An unambiguous $\oplus$-certificate
$\mathcal{U} = \{C_1,\ldots ,C_m\}$ for $f$ is defined to be
a collection of monochromatic \emph{affine subspaces}
that together partition the space $\pmset{n}$.
It is easily seen that a subcube is also an affine subspace.
Analogously, for an unambiguous $\oplus$-certificate $\mathcal{U}$,
on an input $x$, $\UC^\oplus(\mathcal{U},x) := \mathsf{co\text{-}dim}(C_i)$
for the $C_i$ satisfying $x\in C_i$, and
\(\aUC^\oplus(f,\mathcal{U}) := \E_x[\UC^\oplus(\mathcal{U},x)] \).
Similarly, we define $\aUC^\oplus(f)$, $\aUC^\oplus_0(f)$, $\aUC^\oplus_1(f)$, and $\UC^\oplus(f)$. 
\paragraph{DNFs}
A $\dnf$ (disjunctive normal form) is
a disjunction ($\bor$) of conjunctions ($\band$s) of variables and
their negations. An \emph{unambiguous $\dnf$} is a $\dnf$ that
satisfies the additional property that$\colon$
on every $(-1)$-input, exactly one of the conjunctions outputs $-1$. 
Note that an unambiguous $\dnf$ for $f$ is equivalently an unambiguous $1$-certificate for $f$.
\paragraph{Approximate degree}
The $\varepsilon$-\emph{approximate degree} of
$f\colon\pmset{n}\rightarrow \reals$, denoted ${\deg}_{\varepsilon}(f)$,
is defined to be the minimum degree among all multilinear real polynomials
$p$ such that $|f(x)-p(x)| \leq \varepsilon$ for all $x \in \pmset{n}$.
Usually $\varepsilon$ is chosen to be $1/3$,
but it can be chosen to be any constant in $(0,1)$,
without significantly changing the model. 
\paragraph{Deterministic decision trees}
A deterministic decision tree for
$f\colon\pmset{n}\rightarrow \pmset{}$ is a rooted binary tree
where each node is labelled by $i\in [n]$ and the leaves are labelled
with an output bit $\pmset{}$. On input $x\in \pmset{n}$,
the tree proceeds at a node with label $i$ by evaluating the bit $x_i$
and continuing with the subtree corresponding to the value of $x_i$.
Once a leaf is reached, the tree outputs a bit.
We say that a deterministic decision tree computes $f$
if for all $x \in \pmset{n}$ its output equals $f(x)$. 

A parity-decision tree for $f$ is similar to a deterministic decision tree,
except that each node in the tree is labelled by a subset $S\subseteq [n]$.
On input $x\in \pmset{n}$, the tree proceeds at a node with label $S$
by evaluating the parity of the bits $x_i$ for $i\in S$ and
continues with the subtree corresponding to the value of
$\oplus_{i\in S}x_i$. Note that if the subsets at each node
have size $|S|=1$, then we get the standard deterministic decision tree model.
\paragraph{Randomized decision trees}
A randomized decision tree for $f$ is a probability distribution $\mu$
over deterministic decision trees for $f$.
On input $x$, a decision tree is chosen according to $\mu$,
which is then evaluated on $x$.
The complexity of the randomized tree is
the largest depth among all deterministic trees with non-zero probability
of being sampled according to $\mu$.
One can similarly define a randomized parity-decision tree
as a probability distribution over
deterministic parity-decision trees for $f$.

We say that a randomized decision tree computes $f$ with bounded-error 
if for all $x \in \pmset{n}$ its output equals $f(x)$ with probability
at least $2/3$. $R(f)$ (respectively, $R^\oplus(f)$)
denotes the complexity of the optimal randomized (respectively, parity)
decision tree that computes $f$ with bounded-error,
i.e., errs with probability at most $1/3$.
\paragraph{Conical Juntas} Let $h = \sum_{}w_C C$, a \emph{conical junta}, be a weighted sum over conjunctions $C$ of literals with weights $w_C \geq 0$. We say that a Boolean function $f\colon\boolfn{n}$ is represented by a conical junta $h$ if $g=(1-f)/2=h$. More generally, we will consider (non-negative) weighted sum over affine subspaces. We define the \emph{average conical junta degree} of $h$, denoted $\avgcj^{\oplus}(h)$, equals $\sum_{C}w_C\codim(C)2^{-\codim(C)}$. Then, the \emph{average conical junta degree} of $f$ is $\avgcj^{\oplus}(f) \coloneqq \min_{h\text{ a conical junta}}\avgcj^{\oplus}(h)$ the minimum average conical junta degree over all conical junta representations $h$. Similar measures were studied in \cite{GJ16}.

\begin{theorem}[The Reisz-Thorin Interpolation Theorem {\cite[Theorem 6.27]{Folland-book}}]
    \label{thm:reisz-thorin}
    Let $I$ and $J$ be non-empty finite sets. Further let $T\colon \complex^{I}\to\complex^{J}$ be a linear map, and $\nu_i >0$ for $i\in I$ be such that for every $u\in\complex^{I}$, 
    \begin{align}
        \|Tu\|_1 \leq \sum_{i\in I}\nu_i|u_i|,\quad \text { and }\quad \|Tu\|_2^2 \leq \sum_{i\in I}\nu_i|u_i|^2.
    \end{align}
    Then, for every $1\leq r \leq 2$, 
    \begin{align}
       \|Tu\|_r^r =\sum_{j\in J}|(Tu)_j|^r \leq \sum_{i\in I}\nu_i |u_i|^r.
    \end{align}
\end{theorem}

\section{Approximants}
\label{sec:approximants}

\begin{theorem}
    \label{thm:ent-approximants}
    Let $f\colon\cfn{n}$ with $\|f\|_2^2 \leq 1$. Suppose that $a\geq 0$, $C>0$, $L\geq 1$, and for any integer $j \geq 1$, there is a function $P_j\colon\cfn{n}$ such that 
    \begin{align}
    \label{eq:approximant-hypo}
        \|f-P_j\|_2^2 \leq 2^{-j}\quad \text{and}\quad \log\max\{1,\|\widehat{P_j}\|_1\} \leq CL(j+1)^a.  
    \end{align}
    Then, 
    \begin{align}
    \label{eq:approximant-bound}
        \ente(\widehat{f}^2) \leq 6 + 4CK_aL = O_{a,C}(L),
    \end{align}
    where $K_a=\sum_{j\geq 1}2^{-j}(j+3)^a = O(2^aa!)$. 
\end{theorem}
\begin{proof}
    For each $j \geq 1$, define the following set of \emph{good} monomials 
    \begin{align}
    \label{eq:approx-good}
        G_j = \left\{S \mid \fc{f}{S}\neq 0, |\fc{f}{S} - \fc{P_{j+2}}{S}| \leq |\fc{f}{S}|/2 \right\}.
    \end{align}  
    Using $G_j$'s we partition the set of non-zero Fourier coefficients as follows: for each $j\geq 1$,
    \begin{align}
        T_j := G_j\setminus \cup_{k<j}G_k. 
    \end{align}
    Since $\|f-P_j\|_2^2 \to 0$, as $j\to\infty$, every fixed non-zero coefficient eventually becomes good, and thus $T_j$'s partitions the set of non-zero Fourier coefficients of $f$. 
    
    Let us define the Fourier weight $\mu_j$ on the part $T_j$, i.e., $\mu_j := \sum_{S\in T_j}\fcsq{f}{S}$. Also define $B_j :=\max\{1, \|\widehat{P_{j+2}}\|_1\}$ and $\rho := \|f\|_2^2 \leq 1$. 
    From the definitions it follows that $\sum_{j\geq 1}\mu_j = \rho$ and 
    \begin{align}
    \label{eq:approx-1}
       \sum_{k > j}\mu_k \leq \sum_{S \notin G_j} \fcsq{f}{S} \leq \sum_{S \notin G_j}4|\fc{f}{S} - \fc{P_{j+2}}{S}|^2  \leq 4 \|f-P_{j+2}\|_2^2 \leq 2^{-j}, 
    \end{align} 
    where the last inequality follows from \eqref{eq:approximant-hypo}. Further, we have  
    \begin{align}
    \label{eq:approx-2}
    \mu_j = \sum_{S\in T_j}\fcsq{f}{S} \leq \sum_{S\notin G_{j-1}}\fcsq{f}{S} \leq 2^{-(j-1)}.
    \end{align}
    We also have for $S \in T_j$, $|\fc{f}{S}| \leq 2|\fc{P_{j+2}}{S}|$, and therefore, $\sum_{S\in T_j}|\fc{f}{S}| \leq 2B_j$. 

    We can thus bound $\ente(\widehat{f}^2)$ as follows
    \begin{align}
        \ente(\widehat{f}^2) & = \sum_{\substack{j \geq 1 \colon \\ T_j \neq \emptyset}}\sum_{S \in T_j}\fcsq{f}{S}\log \frac{1}{\fcsq{f}{S}} = \sum_{\substack{j \geq 1 \colon \\ T_j \neq \emptyset}}2\mu_j\sum_{S \in T_j}\frac{\fcsq{f}{S}}{\mu_j}\log \frac{1}{|\fc{f}{S}|}, \nonumber \\ 
       & \leq \sum_{j \colon T_j \neq \emptyset} 2\mu_j \log \left(\sum_{S \in T_j}\frac{|\fc{f}{S}|}{\mu_j} \right) \leq 2 \sum_{j\colon T_j \neq \emptyset} \mu_j \log \left(\frac{2B_j}{\mu_j} \right), \nonumber \\
       & = 2\sum_{j} \mu_j \log \frac{2}{\mu_j} + 2\sum_{j} \mu_j\log B_j, \nonumber \\
       & = 2\left( \rho + \sum_{j}j\mu_j + \sum_{j}\mu_j\log \frac{2^{-j}}{\mu_j} \right) + 2 \sum_{j}\mu_j\log B_j, \label{eq:approx-3}
    \end{align}
    where the first inequality follows from the Jensen's inequality using the concavity of log. We now bound each term in \eqref{eq:approx-3} to obtain the bound in \eqref{eq:approximant-bound}. We have 
    \begin{align*}
        \sum_{j\geq 1} j \mu_j & = \sum_{j\geq 1}\mu_j + \sum_{j\geq 1}\sum_{k> j}\mu_k \leq \rho + 1, \\
        \sum_{j\geq 1}\mu_j\log \frac{2^{-j}}{\mu_j} & = \rho \sum_{j\geq 1}\frac{\mu_j}{\rho}\log \frac{2^{-j}}{\mu_j} \leq \rho \log \left( \sum_{\substack{j\geq 1:\\ \mu_j > 0}} \frac{2^{-j}}{\rho} \right) \leq \rho\log\frac{1}{\rho} \leq 2(1-\rho), \text{ and } \\
        \sum_{j\geq 1}\mu_j\log B_j & \leq CL\sum_{j\geq 1}\mu_j(j+3)^a \leq 2CL\sum_{j\geq 1}2^{-j}(j+3)^a = 2CLK_a, 
    \end{align*}
    where we used equations \eqref{eq:approx-1}, \eqref{eq:approx-2}, \eqref{eq:approximant-hypo} and the Jensen's inequality to simplify. Substituting them back in \eqref{eq:approx-3} we obtain
    \begin{align*}
        \ente(\widehat{f}^2) \leq 6 + 4CLK_a. 
    \end{align*} 
\end{proof}
As an application we show that Fourier-entropy is bounded by the logarithm of $\varepsilon$-approximate spectral norm. 

\begin{theorem}
\label{thm:approx-L1-norm}
    Let $f\colon\boolfn{n}$ be a Boolean function. Then, for $\varepsilon > 0$, $\ent(f) = O(\log(1+ \|\widehat{f}\|_{1,\varepsilon}))$.  
\end{theorem}
\begin{proof}
    Without loss of generality, assume $\varepsilon = 1/3$. Let $P'$ be a $1/3$-approximating polynomial for $f$, and let $P := 3P'/4$. Let $\maj_t$ be the Majority function on $t$ variables. Then, define $P_t := \maj_t(P,\ldots ,P)$ and $B := \|\widehat{P}\|_1 \leq \|\widehat{P'}\|_1$. It is well known that the following facts holds:
    \begin{enumerate}
        \item  $\|f-P_t\|_{\infty} \leq 2e^{-t/8}$, and hence $\|f-P_t\|_{2}^2 \leq 4e^{-t/4}$, while   
        \item $\|\widehat{P_t}\|_1 \leq (1+B^2)^{t/2}$. 
    \end{enumerate}
    For each $j\geq 1$, define the approximant $Q_j := P_{4j+8}$. We, therefore, have 
    \begin{align*}
        \|f-Q_j\|_{2}^2 \leq 2^{-j},\quad \text{ and }\quad \log\max\{1, \|\widehat{Q_j}\|_1\} \leq O(j\log (1+B)).  
    \end{align*}
    It, thus follows from Theorem~\ref{thm:ent-approximants}, 
    \begin{align*}
        \ent(\widehat{f}^2) = O(\log(1+B)),\quad \text{ and therefore},\quad  \ent(\widehat{f}^2) = O(\log(1+\|\widehat{f}\|_{1,1/3})). 
    \end{align*}
    The second equality follows since $P'$ is an arbitrary $1/3$-approximating polynomial. 
\end{proof}
As a further corollary to Theorem~\ref{thm:approx-L1-norm}, we bound the Fourier-entropy by the randomized decision-tree complexity. 

\begin{theorem}
    \label{thm:rdt}
    Let $f\colon\boolfn{n}$ be a Boolean function and $\rdt^{\oplus}(f)$ be the randomized parity-decision tree complexity of $f$ with error at most $1/3$. 
    Then, $\ent(\widehat{f}^2) = O(\rdt^{\oplus}(f))$. 
\end{theorem}
\begin{proof}
    Let $\mathcal{T}$ be a randomized $\oplus$-decision tree computing $f$. That is, it is a distribution $\mu$ on deterministic $\oplus$-decision trees $T_1,\ldots , T_t$ such that for each input $x$, $\Pr_{T\sim \mu}[T(x)=f(x)] \geq 2/3$. 
    
    Then, consider the polynomial $P(x)=\sum_{i=1}^t\mu(T_i)p_i(x)$, where $p_i(x)$ are polynomials representing $T_i$ exactly.  Therefore, for each input $x$, $P(x) =\E_{T\sim\mu}[T(x)]$ and thus $|f(x)-P(x)| \leq 2/3$. Hence, $P$ is a $2/3$-approximating polynomial for $f$. Furthermore, $\|\widehat{P}\|_1 \leq 2^{\rdt^{\oplus}(f)}$. The bound now follows from an application of Theorem~\ref{thm:approx-L1-norm}. 
\end{proof}

\section{Moment Inequality}
\label{sec:moment-ineq}

\begin{theorem}
    \label{thm:moment-ineq-orthogonal-sums}
    Let $f\colon \cfn{n}$ and $f=\sum_{i=1}^{m}g_i$ where the non-zero functions $g_i\colon\cfn{n}$ are pairwise orthogonal. Define $p_i := \|g_i\|_2^2$ and $L_i := \|\widehat{g_i}\|_1$. Then, for every $1\leq r \leq 2$, 
    \begin{align}
        \sum_{S\subseteq [n]}|\fc{f}{S}|^r \leq \sum_{i=1}^mL_i^{2-r}p_i^{r-1}. 
    \end{align}
    Equality holds at $r=2$. 
\end{theorem}
\begin{proof}
    Define $\nu_i = L_i^2/p_i$ for $i\in[m]$, and the linear operator $T\colon \complex^m \to\complex^{2^{[n]}}$ by 
    \begin{align}
        Tu = \sum_{i=1}^m \frac{L_i}{p_i}u_i\widehat{g_i}.  
    \end{align}
    We now compute the $\ell_1$ and $\ell_2$ norms of $Tu$. We have, by the triangle inequality, 
    \[\|Tu\|_1 \leq \sum_{i=1}^m\frac{L_i}{p_i}|u_i|L_i = \sum_{i=1}^m\nu_i|u_i|.\]
    The Orthogonality and Parseval give equality for $\ell_2$ norm. That is, 
    \begin{align}
    \label{eq:l2-norm}
        \|Tu\|_2^2=\sum_{i=1}^m\frac{L_i^2}{p_i^2}|u_i|^2\|g_i\|_2^2=\sum_{i=1}^m\nu_i|u_i|^2.
    \end{align}
    Then applying the Reisz-Thorin interpolation theorem (Theorem~\ref{thm:reisz-thorin}), for $1\leq r \leq 2$, we obtain, 
    \[\|Tu\|_r^r \leq \sum_{i=1}^m\nu_i|u_i|^r.\]
    Now choose $u_i=p_i/L_i$ to obtain $Tu=\widehat{f}$, and thereby we get  
    \[ \|Tu\|_r^r = \sum_{S}|\fc{f}{S}|^r \leq \sum_{i=1}^m\nu_i|u_i|^r = \sum_{i=1}^m\nu_i\left(\frac{p_i}{L_i}\right)^r = \sum_{i=1}^mL_i^{2-r}p_i^{r-1}. \]
    Moreover, it is easily seen from \eqref{eq:l2-norm} that equality holds at $r=2$. 
\end{proof}
We now transfer the moment inequality in Theorem~\ref{thm:moment-ineq-orthogonal-sums} to bounds on entropy. 
\begin{theorem}
    \label{thm:ent-ineq-ortho}
     Let $f\colon \cfn{n}$ and $f=\sum_{i=1}^{m}g_i$ where the non-zero functions $g_i\colon\cfn{n}$ are pairwise orthogonal. Define $p_i := \|g_i\|_2^2$ and $L_i := \|\widehat{g_i}\|_1$. Then, 
     \begin{align}
         \ente(\widehat{f}^2) = \sum_{S\subseteq [n]}\fcsq{f}{S}\log\frac{1}{\fcsq{f}{S}}  \leq 2 \sum_{i=1}^mp_i\log\frac{L_i}{p_i}.  
     \end{align}
\end{theorem}
\begin{proof}
    Define the functions 
    \begin{align*}
        F(r) := \sum_{S\subseteq [n]}|\fc{f}{S}|^r,\quad \text{and}\quad G(r) := \sum_{i=1}^mL_i^{2-r}p_i^{r-1}. 
    \end{align*} 
    We have from Theorem~\ref{thm:moment-ineq-orthogonal-sums} that $F(r) \leq G(r)$ for $1\leq r \leq 2$ and $F(2)=G(2)$. Thus for $t>0$, 
    \begin{align}
        \frac{F(2)-F(2-t)}{t} \geq \frac{G(2)-G(2-t)}{t}. 
    \end{align}
    Hence, taking the left-derivative at $2$, gives $F'(2) \geq G'(2)$. Computing we have 
    \begin{align*}
        F'(2) = \sum_{S\subseteq [n]}\fcsq{f}{S}\ln|\fc{f}{S}|,\quad \text{and}\quad G'(2) = \sum_{i=1}^mp_i\ln\frac{p_i}{L_i}.
    \end{align*}
    Multiplying each derivative by $-2/\ln 2$ and using the inequality $F'(2) \geq G'(2)$, we obtain 
    \begin{align}
        \sum_{S\subseteq [n]}\fcsq{f}{S}\log\frac{1}{\fcsq{f}{S}}  \leq 2 \sum_{i=1}^mp_i\log\frac{L_i}{p_i}.
    \end{align}
\end{proof}
As an application of Theorem~\ref{thm:ent-ineq-ortho}, we bound the Fourier-entropy $\ent(\widehat{f}^2)$ by $O(\min\{\aUC_0(f),\aUC_1(f)\})$, the one-sided version of the average unambiguous certificate complexity. 
\begin{theorem}
    \label{thm:ent-auc}
    Let $f\colon\boolfn{n}$ be a Boolean function. Then, $\ent(\widehat{f}^2) = O(\min\{\aUC_0^{\oplus}(f),\aUC_1^{\oplus}(f)\})$, where for $b\in\bool$, $\aUC_b^{\oplus}(f)$ is the average unambiguous $b$-parity-certificate complexity of $f$. 
\end{theorem}
\begin{proof}
    Fix $b=1$ (the other case is analogous). Let $g=(1-f)/2$ and $\mathcal{U}_1=\{C_1,\ldots ,C_m\}$ be an unambiguous $1$-parity-certificate for $f$. Let $g_i\colon\bfn{n}$ be the indicator function of the $1$-parity-certificate $C_i$. Then, it follows by unambiguity that $g=\sum_{i=1}^mg_i$ and $g_i$'s are pairwise orthogonal. From, Lemma~\ref{lem:ente-to-ent} and Theorem~\ref{thm:ent-ineq-ortho}, we have 
    \begin{align}
        \ent(\widehat{f}^2) \leq 4 \ente(\widehat{g}^2) \leq 8 \sum_{i=1}^m\|g_i\|_2^2\log\frac{\|\widehat{g_i}\|_1}{\|g_i\|_2^2}= 8\sum_{i=1}^m\codim(C_i)2^{-\codim(C_i)}=O(\aUC^{\oplus}(f,\mathcal{U}_1)). 
    \end{align}
    Since $\mathcal{U}_1$ is arbitrary, we obtain $\ent(\widehat{f}^2) = O(\aUC_1^{\oplus}(f))$.
\end{proof}
We now generalize Theorem~\ref{thm:moment-ineq-orthogonal-sums} (and Theorem~\ref{thm:ent-ineq-ortho}) to handle the case of conical juntas where the conjunctions may not be orthogonal. 

\begin{theorem}
    \label{thm:moment-ineq-conical-junta}
    Let $f\colon \cfn{n}$ and $f=\sum_{i=1}^{m}g_i$ where $g_i\colon\cfn{n}$'s are non-zero functions. Define $m\times m$ Gram matrix $\Gamma$ where $\Gamma_{ij} := \langle g_i,g_j\rangle=\E[\overline{g_i}g_j]$, and denote $L_i := \|\widehat{g_i}\|_1$. Suppose there are numbers $a_i>0$ such that 
    \begin{align}
        \label{eq:assumption-conical-junta}
        \Gamma \preceq D \coloneqq \mathsf{diag}(a_1,\ldots ,a_m). 
    \end{align}
    Then, for every $1\leq r \leq 2$,
    \begin{align}
        \sum_{S\subseteq [n]}|\fc{f}{S}|^r \leq \sum_{i=1}^mL_i^{2-r}a_i^{r-1}. 
    \end{align} 
    If, additionally, the equality holds at $r=2$, $ \sum_{i}a_i=\|f\|_2^2$, then 
    \begin{align}
        \ente(\widehat{f}^2) \leq 2 \sum_{i=1}^m a_i\log\frac{L_i}{a_i}. 
    \end{align}
\end{theorem}
\begin{proof}
    Define $\nu_i = L_i^2/a_i$ for $i\in[m]$, and the linear operator $T\colon \complex^m \to\complex^{2^{[n]}}$ by 
    \begin{align}
        Tu = \sum_{i=1}^m \frac{L_i}{a_i}u_i\widehat{g_i},\quad \text{for all }u\in \complex^m.   
    \end{align}
    As before we compute the $\ell_1$ and $\ell_2$ norms of $Tu$. We have, by the triangle inequality, 
    \[\|Tu\|_1 \leq \sum_{i=1}^m\frac{L_i}{a_i}|u_i|\|\widehat{g_i}\|_1 = \sum_{i=1}^m\frac{L_i^2}{a_i}|u_i|= \sum_{i=1}^m\nu_i|u_i|.\]
    For $\ell_2$ norm, we first use Parseval and then \eqref{eq:assumption-conical-junta}. That is, 
    \begin{align*}
        \|Tu\|_2^2=\left\|\sum_{i=1}^m \frac{L_i}{a_i}u_i\widehat{g_i}\right\|_2^2 = \left\|\sum_{i=1}^m \frac{L_i}{a_i}u_ig_i\right\|_2^2 \leq \sum_{i=1}^ma_i\left|\frac{L_i}{a_i}u_i\right|^2 \leq \sum_{i=1}^m\nu_i|u_i|^2,
    \end{align*}
    where the first inequality follows from \eqref{eq:assumption-conical-junta} which implies that for all $c\in\complex^m$, $\|\sum_{i=1}^m c_ig_i\|_2^2 \leq \sum_{i=1}^ma_i|c_i|^2$. 
    Then applying the Reisz-Thorin interpolation theorem (Theorem~\ref{thm:reisz-thorin}), for $1\leq r \leq 2$, we obtain, 
    \[\|Tu\|_r^r \leq \sum_{i=1}^m\nu_i|u_i|^r.\]
    Now choose $u_i=a_i/L_i$ to obtain $Tu=\widehat{f}$, and thereby we get  
    \[ \|Tu\|_r^r = \sum_{S}|\fc{f}{S}|^r \leq \sum_{i=1}^m\nu_i|u_i|^r = \sum_{i=1}^m\nu_i\left(\frac{a_i}{L_i}\right)^r = \sum_{i=1}^mL_i^{2-r}a_i^{r-1}. \]
    Furthermore, if $\sum_{i}a_i=\|f\|_2^2$, then the arguments from the proof of Theorem~\ref{thm:ent-ineq-ortho} can be applied here too to obtain a similar bound on entropy 
    \begin{align*}
        \ente(\widehat{f}^2) \leq 2\sum_{i=1}^ma_i\log\frac{L_i}{a_i}. 
    \end{align*}
\end{proof}
We now use Theorem~\ref{thm:moment-ineq-conical-junta} to bound Fourier-entropy by the average conical junta degree. This measure is the average of the measure 'average degree' defined and studied by \cite{GJ16}. 

\begin{theorem}
\label{thm:ent-conical-junta}
    Let $f\colon\boolfn{n}$ be a Boolean function. Then, 
    \begin{align*}
    \ent(\widehat{f}^2) = O(\min\{\avgcj^{\oplus}(f),\avgcj^{\oplus}(-f)\}).
    \end{align*}
\end{theorem}
\begin{proof}
    Let $f$ be represented by a conical junta $h=\sum_{i=1}^mw_iC_i$ where $C_i$'s are conjunctions of literals (or, more generally, indicators of affine subspaces). Recall, $\avgcj(h)=\sum_{i=1}^mw_i\codim(C_i)2^{-\codim(C_i)}$. Since Fourier-entropy is invariant under negation, it suffices to show $\ent(\widehat{f}^2) = O(\avgcj(h))$. 

    Let $g=(1-f)/2=h=\sum_{i=1}^mg_i$, where $g_i = w_iC_i$. Furthermore, let $\Gamma$ be a $m\times m$ Gram matrix with $\Gamma_{ij}=\langle g_i,g_j\rangle$. Then, $\Gamma_{i,j}\geq 0$ and define the row sums $a_i \coloneqq \sum_{j}\Gamma_{ij}=\langle g_i,g\rangle$ and the diagonal matrix $D \coloneqq \mathsf{diag}(a_1,\ldots ,a_m)$. It is then easily seen that $\Gamma \preceq D$. Indeed, for every $c\in\complex^m$, 
    \begin{align*}
        \|\sum_i c_ig_i \|_2^2 - \sum_{i}a_i|c_i|^2 = -\frac{1}{2}\sum_{i,j}\Gamma_{ij}|c_i-c_j|^2 \leq 0. 
    \end{align*}
    Also $\sum_{i}a_i=\sum_{i}\langle g_i,g\rangle = \|g\|_2^2$, $L_i := \|\widehat{g_i}\|_1 = w_i$, and $a_i = w_i2^{-\codim(C_i)}$. Thus, by an application of Lemma~\ref{lem:ente-to-ent} and Theorem~\ref{thm:moment-ineq-conical-junta}, we obtain 
    \begin{align*}
        \ent(\widehat{f}^2) = O(\ente(\widehat{g}^2)) = O\left(\sum_{i}w_i2^{-\codim(C_i)}\codim(C_i)\right)= O(\avgcj(h)). 
    \end{align*}
\end{proof}

\section{Application to Quantum Communication Complexity Lower Bound}
\label{sec:application-qcc}
Let $X,Y$ be finite non-empty sets and 
let $M\in\{-1,1\}^{X\times Y}$ be a total sign matrix. The quantity
$\Qcc_\varepsilon(M)$ is the minimum worst-case number of qubits communicated by a
two-party protocol that outputs $M_{xy}$ with error at most $\varepsilon$ on every
input $(x,y)$. Arbitrary finite-dimensional prior entanglement and local
ancillas are allowed. Bob is the designated output party. 
For probability vectors $u=(u_x)_{x\in X}$ and $v=(v_y)_{y\in Y}$, let
\begin{align}
\label{eq:weighted-matrix}
 A_{u,v}=D_{\sqrt u}MD_{\sqrt v},\quad \text{where }
 D_{\sqrt u}=\mathsf{diag}(\sqrt{u_x})_{x\in X}\,\text{ and }
 D_{\sqrt v}=\mathsf{diag}(\sqrt{v_y})_{y\in Y}.
\end{align}
Let $\sigma_1(A_{u,v})\ge\sigma_2(A_{u,v})\ge\cdots\ge0$ be the singular
values of $A_{u,v}$. Since
\begin{align}
\label{eq:normalization}
 \sum_i\sigma_i(A_{u,v})^2=\norm{A_{u,v}}_{\Frob}^2 =\sum_{x,y}u_xv_y|M_{xy}|^2=1,
\end{align}
their squares form a probability distribution. Define
\begin{equation}
 H_{u,v}(M)=\sum_i\sigma_i(A_{u,v})^2
                   \log\frac{1}{\sigma_i(A_{u,v})^2},
 \label{eq:entropy}
\end{equation}
with $0\log(1/0)=0$.
We now show that supremum over all probability distributions $u$ and $v$ lower bounds the quantum communication complexity. 
\begin{theorem}
\label{thm:ent-qcc}
For every $0\le\varepsilon<1/2$, there is a constant $c_\varepsilon>0$ such that,
for every total sign matrix $M$ and every pair of probability vectors $u,v$,
\begin{align*}
 \Qcc_\varepsilon(M)\ge c_\varepsilon H_{u,v}(M).
\end{align*}
In particular, for $f\colon\boolfn{n}$, let $M_f(x,y)=f(x\oplus y)$, $u$ and $v$ be uniform distributions, and then we have
\begin{align*}
    \Qcc_{\varepsilon}(M_f) \geq c_\varepsilon \ent(\widehat{f}^2). 
\end{align*}
\end{theorem}
For uniform $N\times N$ sign matrices, let $E =H_{u,v}(M)$. Linial--Shraibman prove the bounded-error estimate 
\begin{align*}
 \Qcc_{1/3}(M)\ge
 \Omega\!\left(\frac{E}{1+\log((\log N)/E)}\right),\qquad E>0,
\end{align*}
in \cite[Theorem~30]{LS07}. Theorem~\ref{thm:ent-qcc} removes the denominator and provides a dimension-independent bound. We recall some basics of matrix norms. For a matrix $B$, let $\norm B_{\tn}=\sum_i\sigma_i(B)$ denote its trace
norm, and let $\norm B_{\max}=\max_{x,y}|B_{xy}|$.
Define the factorization norm
\begin{align}
\label{eq:gamma}
 \gamma_2(B)=\inf_{B=UV^*}
       \left(\max_x\norm{U_x}_2\right)
       \left(\max_y\norm{V_y}_2\right),
\end{align}
where $U_x,V_y$ are rows. We need the following facts about various norms.
\begin{lemma}[{\cite[Corollary~IV.2.6]{Bhatia97}}]
    \label{lem:trace-gamma-2}
    For every matrix $B$ and every pair of probability vectors $u,v$,
    \begin{align*}
    \norm{D_{\sqrt  u}BD_{\sqrt v}}_{\tn}\le\gamma_2(B).
    \end{align*}
\end{lemma}
\begin{lemma}[{\cite[Theorem~IV.3.4]{Bhatia97}}]
    \label{lem:singular-F-norm}
    If $A,B$ have the same dimensions and $a_i,b_i$ are their singular values in decreasing order, then
\begin{align*}
 \sum_i(a_i-b_i)^2\le\norm{A-B}_{\Frob}^2.
\end{align*}

\end{lemma}
We also need the following lemma from \cite{LS07} about $\gamma_2$-norm of the matrix of acceptance probabilities.  
\begin{lemma}[{\cite[Lemma 10]{LS07}}]
    \label{lem:gamma-2-qcc}
    If $P_{xy}$ is the acceptance probability of a protocol with communication cost $q$, then \(\gamma_2(P)\le2^q\). 
For its expected signed output matrix $B=2P-J$, where $J$ is all ones,
\begin{align*}
 \gamma_2(B)\le 2\gamma_2(P) + 1\le2^{q+1}+1\le2^{q+2}.
\end{align*}
\end{lemma}
We are now ready to prove Theorem~\ref{thm:ent-qcc}. 
\begin{proof}[Proof of Theorem~\ref{thm:ent-qcc}]
Let $M$ be any sign matrix, and assume, without loss of generality, $q:=\Qcc_\varepsilon(M) \geq 1$. For every integer \(j\ge1\), amplify a cost \(q\) protocol by independent repetition and majority until its error is at most \(\eta_j=2^{-j/2-1}\). Let the cost of this amplified protocol be $q_j =O_{\varepsilon}(q(j+1))$. Let \(P_j(x,y)\) be the probability that this amplified protocol outputs \(+1\), and define its expected signed output
\(B_j=2P_j-J\), where \(J\) is the all-ones matrix. From the correctness of the protocol, we then have 
\begin{align}
\label{eq:matrix-approximants}
 \norm{M-B_j}_{\max}\le2\eta_j=2^{-j/2},\quad \text{ and }\quad
 \gamma_2(B_j)\le2^{q_j+2},
\end{align}
where the second estimate is Lemma~\ref{lem:gamma-2-qcc}.
Let $A=A_{u,v}$ and $A_j=D_{\sqrt u}B_jD_{\sqrt v}$. Then pointwise
approximation \eqref{eq:matrix-approximants} imply
\begin{align}
 \norm{A-A_j}_{\Frob}^2
 =\sum_{x,y}u_xv_y|M_{xy}-(B_j)_{xy}|^2
 \le2^{-j}.
 \label{eq:frob-approx}
\end{align}
Lemma~\ref{lem:trace-gamma-2} gives 
\(\norm{A_j}_{\tn}\le\gamma_2(B_j)\le2^{q_j+2}\), and therefore $\log\max\{1,\norm{A_j}_{\tn}\} = O_\varepsilon(q(j+1))$. 
Let
\[ s=(\sigma_i(A))_i,\qquad s^{(j)}=(\sigma_i(A_j))_i,\]
with singular values listed in decreasing order and zeroes included so all lists have the same length. We therefore have, from Lemma~\ref{lem:singular-F-norm} and \eqref{eq:frob-approx},
\begin{align*}
    \norm{s-s^{(j)}}_2^2\le2^{-j}.
\end{align*}
Moreover, $\norm{s}_2=1$ and $\norm{s^{(j)}}_1=\norm{A_j}_{\tn}$, so
\(\log\max\{1,\norm{s^{(j)}}_1\}= O_\varepsilon(q(j+1))\). Now using Theorem~\ref{thm:ent-approximants} with $s$ and $s^{(j)}$'s gives
\begin{align*}
    H_{u,v}(M)=O_{\varepsilon}(q)=O_{\varepsilon}(\Qcc_{\varepsilon}(M)). 
\end{align*}
In particular, when $M_f$ is a XOR-matrix of a Booelean function $f\colon\boolfn{n}$, then Fourier-entropy of $f$ lower bounds the quantum communication cost of $f(x\oplus y)$, 
\[\Qcc_{\varepsilon}(M_f) = \Omega_{\varepsilon}(\ent(\widehat{f}^2)).\]

\end{proof}

\paragraph{AI statement.} We used several ChatGPT and Claude models 
over the last few months to explore proof strategies, search for related literature, and revise the manuscript. 
The authors had a part of the proof idea which was substantially generalized by these AI models. 
The authors examined these arguments, identified and corrected errors, and
significantly simplified proofs and their presentation.
The authors take full responsibility for the correctness and content
of the paper.

\bibliographystyle{alpha}
\bibliography{entropy}

\end{document}